\documentclass[twocolumn,showpacs,prl,superscriptaddress,nobibnotes]{revtex4}
\usepackage{graphicx}

\begin{document}

\title{Synchronization in nonlinear oscillators with conjugate coupling}
\author{Wenchen Han}
\address{School of Science, Beijing University of Posts and
Telecommunications, Beijing, 100876, People's Republic of China}
\author{Mei, Zhang}\email{meizhang@bnu.edu.cn}
\address{Department of Physics, Beijing Normal University, Beijing,
100875, People's Republic of China}
\author{Junzhong Yang}\email{jzyang@bupt.edu.cn} \address{School of
Science, Beijing University of Posts and Telecommunications,
Beijing, 100876, People's Republic of China}

\date{\today}

\begin{abstract}
In this work, we investigate the synchronization in
oscillators with conjugate coupling in which oscillators interact
via dissimilar variables. The synchronous dynamics and its
stability are investigated theoretically and numerically. We find
that the synchronous dynamics and its stability are dependent on
both coupling scheme and the coupling constant. We also find that
the synchronization may be independent of the number of
oscillators. Numerical demonstrations with Lorenz oscillators are
provided.
\end{abstract}


\maketitle The study of synchronization phenomena in coupled
periodic oscillators has been active since the early days of
physics \cite{boc06,are08}. Chaos implies sensitive dependence on
initial conditions, with nearby trajectories diverging
exponentially, and the synchronization among chaotic oscillators
has become a topic of great interest since 1990
\cite{peco90,piko84}. The general theories on complete
synchronization in which the distance between states of
interacting identical chaotic units approaches zero for
$t\rightarrow \infty$ have been well framed \cite{yang98,pec98}.
In these theories, chaotic oscillators interact with each other
through the same (nonconjugate) variables of different
oscillators. However, coupling via dissimilar (conjugate)
variables is also natural in real situations \cite{gol96,uet03}.
One example is the coupled-semiconductor-laser experiments by Kim
and Roy \cite{kim05}, where the photon intensity fluctuation from
one laser is used to modulate the injection current of the other,
and vice versa. In the nonconjugate coupling case, the interaction
term vanishes with the buildup of complete synchronization and the
synchronous state is a solution of isolated system. In contrast,
the interaction term in conjugate coupling case may stay nonzero
even when the units are synchronized.

The dynamical system with conjugate coupling has been used to
realize the amplitude death \cite{bar85,ram98} in coupled
identical units, the phenomenon in which unstable equilibrium in
isolated unit becomes stable with the assistance of the coupling,
in several recent works \cite{kar07,sin11,sha11}. Interestingly,
the realized amplitude death has indeed referred that
synchronization in oscillators with conjugate coupling is possible
but the synchronous state is not necessarily a stable solution of
isolated units. Then questions arise: Can synchronization in
chaotic oscillators with conjugate coupling be realized? What is
the synchronous state in chaotic oscillators with conjugate
coupling and how about its stability?

The main goal in this work is to theoretically investigate the
synchronous dynamics in a ring of identical chaotic oscillators
with conjugate coupling and its stability by following the methods
in Ref.~\cite{yang98,pec98}. The statements are demonstrated
through numerical simulations with Lorenz oscillators. We also
show that the statements are valid for regular random networks in
which each oscillator has the same number of neighbors.

The model we consider takes the general form
\begin{equation}\label{eq_1}
\dot{\mathbf{x}}_{i}=\mathbf{f}(\mathbf{x}_{i})+\epsilon
(\mathcal{D}_{2}\mathbf{x}_{i+1}-\mathcal{D}_{1}\mathbf{x}_{i})
+\epsilon(\mathcal{D}_{2}\mathbf{x}_{i-1}-\mathcal{D}_{1}\mathbf{x}_{i}) \\
\end{equation}
where $\mathbf{x}_{i}\in R^{n} (i=1,2,\cdots,N)$,
$\mathbf{f}:R^{n}\rightarrow R^{n}$ is nonlinear and capable of
exhibiting rich dynamics such as chaos. The periodic boundary
conditions are imposed on Eq.~\ref{eq_1}. The parameter $\epsilon$
is a scalar coupling constant. $\mathcal{D}_{1}$ and
$\mathcal{D}_{2}$ are constant matrices describing coupling
schemes. When $\mathcal{D}_{1}=\mathcal{D}_{2}$, the interaction
terms become
$\mathcal{D}_{1}(\mathbf{x}_{i+1}+\mathbf{x}_{i-1}-2\mathbf{x}_{i})$
and the ordinary non-conjugate coupled oscillators are recovered
in which oscillators interact with each other through the same
variables.

Now we are interested in synchronous states; the states reside on
a synchronization manifold defined by
$M=\{(\mathbf{x}_{1},\cdots,\mathbf{x}_{n}):\mathbf{x}_{i}=\mathbf{s}(t)\}$
where $\mathbf{s}(t)$ satisfies the equation of motion
\begin{eqnarray}\label{eq_2}
\dot{\mathbf{s}}=\mathbf{f}(\mathbf{s})+2\epsilon
(\mathcal{D}_{2}-\mathcal{D}_{1})\mathbf{s}.
\end{eqnarray}
To be noted that the synchronous state is not the solution of the
isolated oscillator any more and its dynamics depends on the
coupling constant and the matrices $\mathcal{D}_{1}$ and
$\mathcal{D}_{2}$ (or the coupling scheme). The stability of the
synchronous state can be investigated by letting
$\mathbf{x}_{i}=\mathbf{s}+\mathbf{\xi}_{i}$ and linearizing
Eq.~(\ref{eq_1}) about $\mathbf{s}(t)$. This leads to
\begin{eqnarray}\label{eq_3}
\frac{d}{dt}\mathbf{\xi}&=&I\otimes(D\mathbf{f}(\mathbf{s})-2\epsilon\mathcal{D}_{1})\mathbf{\xi}
+\epsilon \mathbf{C}\otimes\mathcal{D}_{2}\mathbf{\xi}
\end{eqnarray}
where $D\mathbf{f}(\mathbf{s})$ is the Jacobian of $\mathbf{f}$ on
$\mathbf{s}$, $I$ is the $N\times N$ unit matrix, the coupling
matrix $\mathbf{C}$ is an $N\times N$ matrix with zero elements
except that $c_{i,i+1}=c_{i-1,i}=1$, which describes the
interaction among oscillators. The eigenvalues and eigenvectors of
$C$ satisfy $C\phi_{i}=\lambda_{i}\phi_{i}$. By expanding
$\mathbf{\xi}$ into the eigenvectors of $C$,
$\mathbf{\xi}=\Sigma_{i=1}^{N}\mathbf{\eta}_{i}\mathbf{\phi}_{i}$
where $\mathbf{\eta}_{i}$ is the coefficient and is dependent on
time, the linear stability equations is diagonalized and gives
\begin{eqnarray}\label{eq_4}
\dot{\mathbf{\eta}}_{i}&=&[D\mathbf{f}(\mathbf{x}^{*})-2\epsilon\mathcal{D}_{1}+\epsilon \lambda_{i}\mathcal{D}_{2}]\eta_{i},\nonumber\\
i&=&1,2,\cdots,N,
\end{eqnarray}
where $\lambda_{i}=2\cos\frac{2i\pi}{N}$ is the eigenvalue of $C$.
It can be demonstrated that the synchronous manifold coincides
with the subspace spanned by the eigenvector of $C$ with
eigenvalue $\lambda_{N}=2$. The Lyapunov exponents given by the
linear stability equation with $\lambda_{N}=2$ determine the
dynamics of the synchronous state while the modes characterized by
all other eigenvalues govern the motion transversal to the
synchronous manifold. Suppose that the mode with $\lambda_{i}$
gives Lyapunov exponents $\Lambda^{(i)}_{1}\geq
\Lambda^{(i)}_{2}\geq \cdots\geq \Lambda^{(i)}_{n}$. Then the
stability of the synchronous state requires $\Lambda^{(i)}_{1}<0$
for all $i$ ranging from $1$ to $N-1$.

\begin{figure}
\includegraphics[width=3.4in]{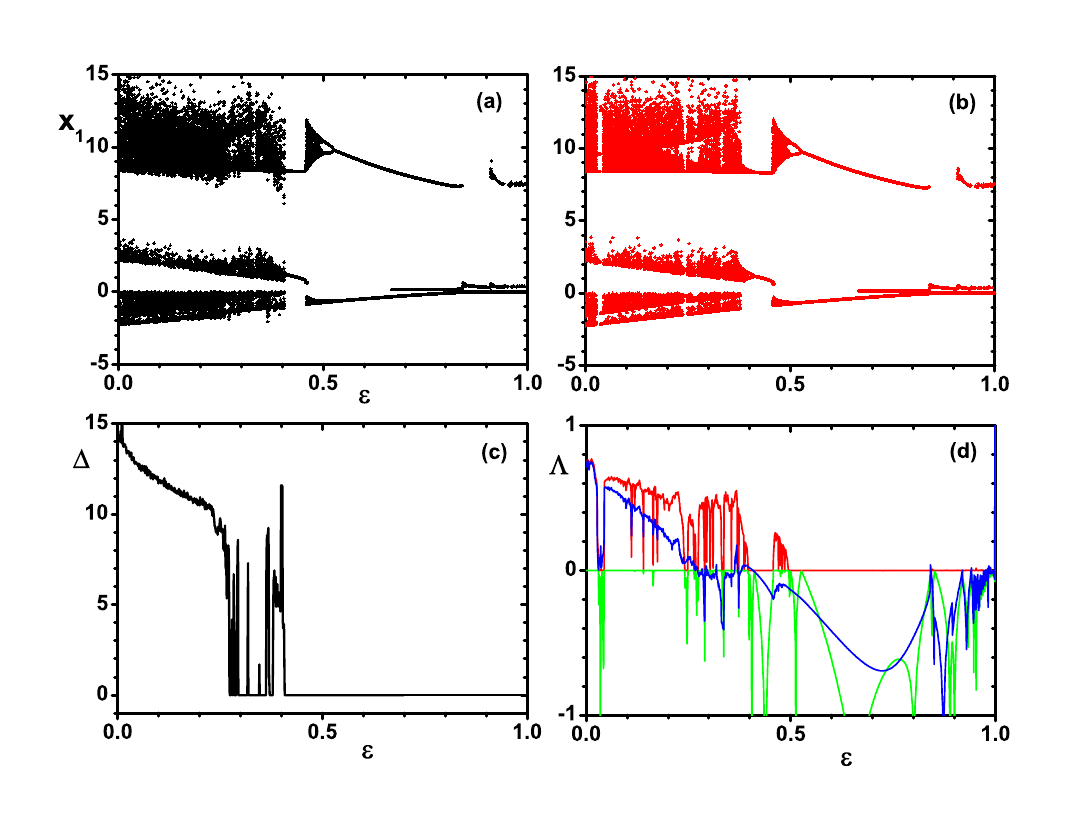}%
\caption{(a) The bifurcation diagram of one
oscillator in a pair of coupled Lorenz oscillators. (b) The
bifurcation diagram of the synchronous motion which follows Eq.
(2) but with $2\epsilon$ replaced by $\epsilon$. (c) The
synchronization error $\Delta$ is plotted against the coupling
constant, which shows that the synchronization error depends on
$\epsilon$ in a non-monotonic way. (d) The first two largest
Lyapunov exponents of the synchronous motion ($\Lambda^{(2)}_{1}$
in red and $\Lambda^{(2)}_{2}$ in green) and the largest Lyapunov
exponent $\Lambda^{(1)}_{1}$ of the transversal mode (in blue) are
plotted against $\epsilon$. $\sigma=10$, $r=28$, and $\beta=1$.
The matrices $D_{1}$ and $D_{2}$ are presented in the text.}
\label{fig_1}
\end{figure}

\begin{figure}
\begin{center}
\includegraphics[width=2in]{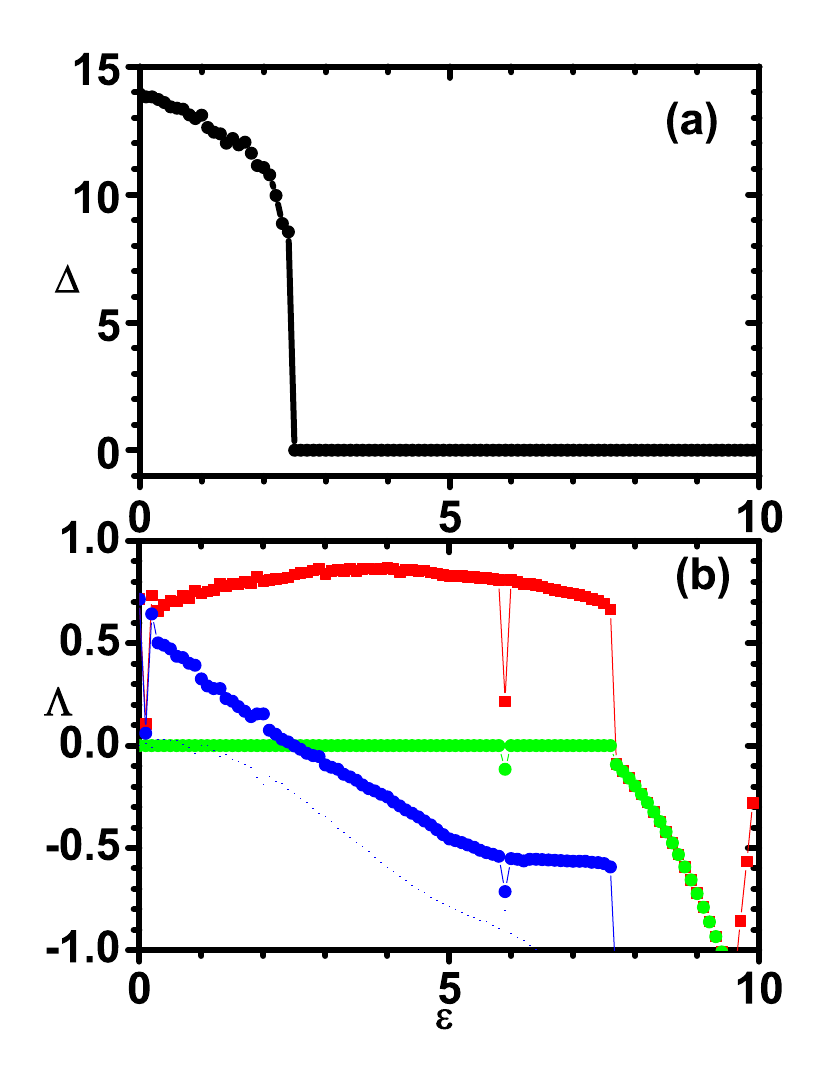}%
\end{center}
\caption{\label{fig_2} The dynamics of a pair of Lorenz
oscillators with $D_{1}$ and $D_{2}$ presented in the text. (a) The
synchronization error $\Delta$ is plotted against the coupling
constant. (b) The first two largest Lyapunov exponents of the
synchronous motion ($\Lambda^{(2)}_{1}$ in red and
$\Lambda^{(2)}_{2}$ in green) and the largest Lyapunov exponent
$\Lambda^{(1)}_{1}$ of the transversal mode (in blue) are plotted
against the coupling constant. $\sigma=10$, $r=28$, and
$\beta=1$.}
\end{figure}


To make above analysis clear, we take some specific systems as
examples. We begin with a pair of oscillators coupled together
which is a special case of a ring. For a pair of oscillators where
$N=2$, there is only one coupling term in Eq.(1), therefore the
motion equation of the synchronous state Eq.(2) and the linear
stability equation Eq.(4) should be modified by replacing the
factor $2\epsilon$ with $\epsilon$. Correspondingly, the
eigenvalues are changed to be $\lambda_{i}=\mp 1, (i=1,2)$ in
which the mode with $\lambda_{2}=1$ accounts for the synchronous
motion. Firstly, we consider two identical Lorenz oscillators
coupled conjugately in which
$\mathcal{D}_{1}=\left(\begin{array}{ccc}0&0&0\\0&0&0\\1&0&0\end{array}\right)$
and
$\mathcal{D}_{2}=\left(\begin{array}{ccc}0&0&0\\0&0&0\\0&0&1\end{array}\right)$.
The motion of a Lorenz oscillators follows
\begin{eqnarray}\label{eq_6}
\dot{x}&=&\sigma (y-x),\nonumber\\
\dot{y}&=&rx-y-xz,\nonumber\\
\dot{z}&=&xy-\beta z.
\end{eqnarray}
Eq.~(\ref{eq_6}) has a chaotic attractor for parameters
$\sigma=10$, $r=28$, and $\beta=1$. Figure \ref{fig_1}(a) and (b)
show the bifurcation diagrams of coupled oscillators and the
synchronous motion against the coupling constant, respectively.
With the variance of the coupling strength, rich dynamics is
found. Especially, the synchronous motion displays various
periodic windows in which the regular motions transit to chaotic
ones. The resemblance between the diagrams of coupled oscillators
and the synchronous motion indicates that there is no other stable
attractors other than the synchronous solution even if the
synchronous motion is unstable. Figure 1(c) shows the
synchronization error
$\Delta=\langle\sqrt{(x_{1}-x_{2})^2+(y_{1}-y_{2})^2+(z_{1}-z_{2})^2}\rangle$
in which $\langle\cdot\rangle$ means the average over a long time
interval after transient. $\Delta$ is a measure on
desynchronization and the two oscillators get synchronized when
$\Delta=0$. The dependence of $\Delta$ on $\epsilon$ in Fig.1(c)
reveals an interesting feature that there exists one regime of
$\epsilon$ in which the synchronization and desynchronization
alternate strongly, which is not quite common. Figure 1(d) shows
the first two Lyapunov exponents $\Lambda^{(2)}_{1,2}$ of the
synchronous motion and the largest Lyapunov exponent
$\Lambda^{(1)}_{1}$ of the transversal mode against the coupling
strength $\epsilon$. The first two Lyapunov exponents of the
synchronous motion show plenty of periodic windows in which the
period-doubling bifurcation sequence can be found.
$\Lambda^{(1)}_{1}$ stays at zero for $\epsilon>0.41$ and
fluctuates around zero in the range of $\epsilon\in(0.26,0.41)$.
As analyzed above, the negative largest Lyapunov exponent of the
transversal model indicates the synchronization between
oscillators. The stability regime of the synchronous motion
indicated by negative $\Lambda^{(1)}_{1}$ is in agreement with
that shown by $\Delta$ in Fig.1(c). To be addressed, the
synchronization between two oscillators does not rely on whether
the synchronous dynamics is periodic or chaotic as shown in Fig.1.

The second example is still the Lorenz oscillator but with
$\mathcal{D}_{1}=\left(\begin{array}{ccc}0&1&0\\0&0&0\\0&0&0\end{array}\right)$
and
$\mathcal{D}_{2}=\left(\begin{array}{ccc}1&0&0\\0&0&0\\0&0&0\end{array}\right)$.
The transition to the synchronization at $\epsilon=2.5$ can be
found either from $\Delta$ in Fig.2(a) or from the the largest
$\Lambda^{1}_{1}$ of the transversal mode in Fig.2(b). Figure 2(b)
shows that the stable synchronous motion could be chaotic or
time-independent. It can be found that there are two stable
time-independent synchronous solutions which satisfy $x=y\neq0$
and are symmetrical under transformation of
$(x,y,z)\rightarrow(-x,-y,z)$. The properties of the stable
time-independent synchronous solutions indicate that they are
actually the pair of unstable fixed points in the isolated Lorenz
oscillator. In terminology, amplitude death refers to a situation
where individual oscillators cease to oscillate when coupled and
settle down to their unstable equilibrium. Previous studies have
shown that systems with different chaotic oscillators
non-conjugately coupled can give rise to time-independent solution
either. However, these realized solutions do not set the systems
onto unstable equilibria of their own and such phenomena are
always named as phase death \cite{bar83,zey01,zhu08} or quenched
death. It is an interesting finding that using conjugate coupling
may generate amplitude death in coupled chaotic oscillators in its
original sense. The state of amplitude death in Fig.2 loses its
stability with the decrease of the coupling constant $\epsilon$
through a subcritical Hopf bifurcation, which gives rise to a
synchronous chaotic motion.

\begin{figure}
\begin{center}
\includegraphics[width=3in]{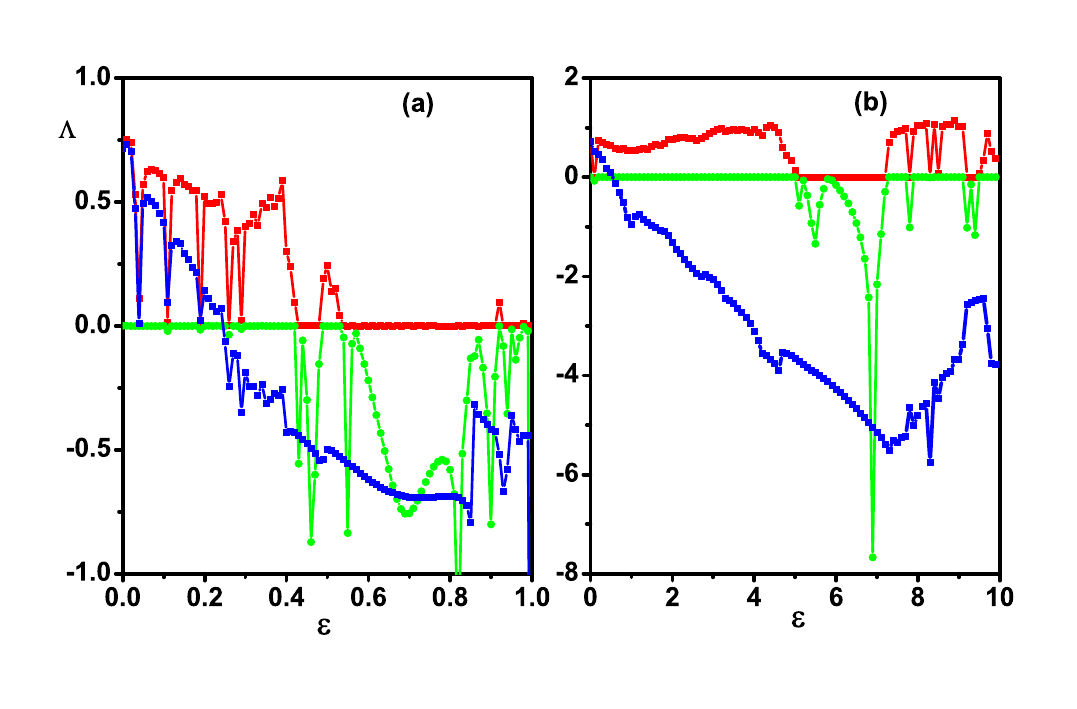}%
\end{center}
\caption{\label{fig_6} The dynamics of a pair of Lorenz
oscillators. The Lyapunov exponents of the synchronous motion
($\Lambda^{(2)}_{1}$ in red and $\Lambda^{(2)}_{2}$ in green) and
the largest Lyapunov exponent $\Lambda^{(1)}_{1}$ of the
transversal mode (in blue) are plotted against the coupling
constant. The matrices $D_{1}$ and $D_{2}$ in plots (a) and (b)
are presented in the text. $\sigma=10$, $r=28$, and $\beta=1$.}
\end{figure}

There is only one nonzero element in the matrices $\mathbf{D}_{1}$
and $\mathbf{D}_{2}$ in the above two examples. However, the
coupling schemes in Eq.(1) may be more complicated than them. For
example,
$\mathcal{D}_{1}=\left(\begin{array}{ccc}0&1&0\\0&0&0\\1&0&0\end{array}\right)$
and
$\mathcal{D}_{2}=\left(\begin{array}{ccc}1&0&0\\0&0&0\\0&0&1\end{array}\right)$
in Fig.~\ref{fig_6}(a), and
$\mathcal{D}_{1}=\left(\begin{array}{ccc}1&0&0\\1&0&0\\1&0&0\end{array}\right)$
and
$\mathcal{D}_{2}=\left(\begin{array}{ccc}0&1&0\\0&1&0\\0&1&0\end{array}\right)$
in Fig.~\ref{fig_6}(b). In these two examples, there are two and
three non-zero elements in the matrices $\mathbf{D}_{1}$ and
$\mathbf{D}_{2}$, respectively. In each case, we find that
coupling strength may adjust the system from desynchronization to
synchronization. It is worth mentioning that the oscillators with
conjugate coupling may provide plenty of coupling schemes. For
example, in coupled Lorenz oscillators, there are eighteen
coupling schemes even if there is only one non-zero elements in
$\mathbf{D}_{1}$ and $\mathbf{D}_{2}$. Large number of coupling
schemes may give rise to rich synchronous dynamics and
de-synchronous dynamics, which may render the oscillators with
conjugate coupling a platform for investigating exotic nonlinear
dynamics and pattern formation.

\begin{figure}
\includegraphics[width=3.4in]{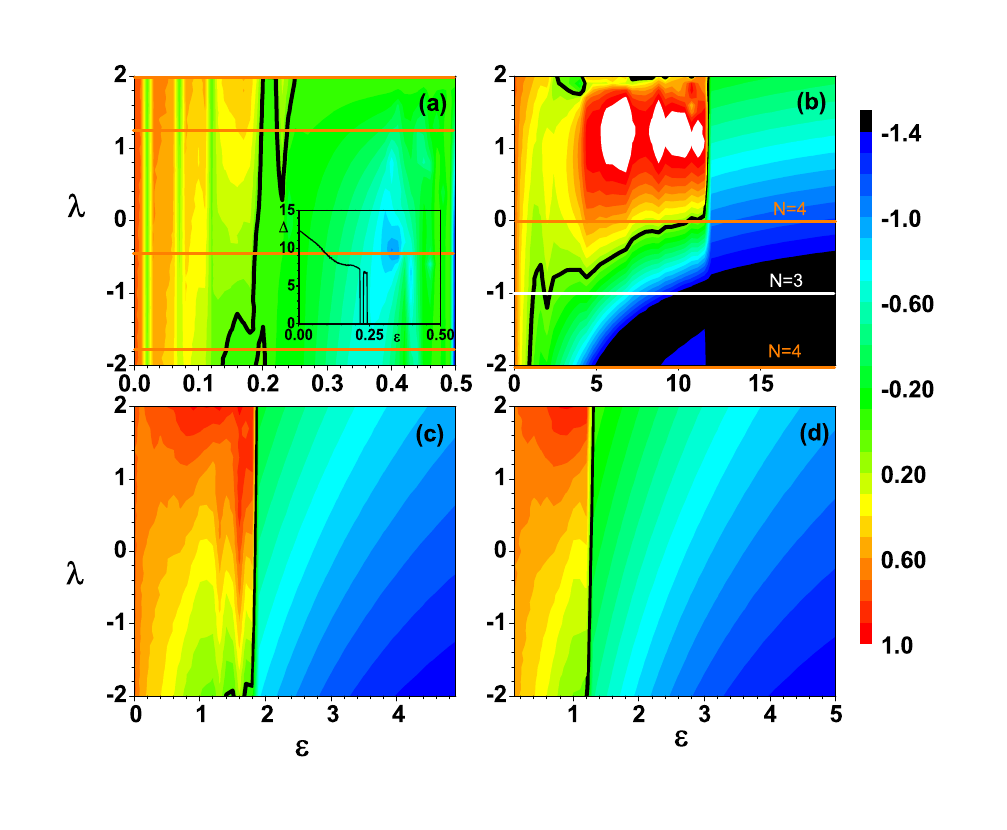}%
\caption{\label{fig_4} The dynamics of a ring of Lorenz
oscillators with $\sigma=10$, $r=28$, and $\beta=1$. The
dependence of the largest Lyapunov exponent $\Lambda_{1}$ on the
coupling constant $\epsilon$ and the parameter $\lambda$ for
different coupling schemes (see the text). The black curves in the
plots denote the set of $\lambda$ and $\epsilon$ at which
$\Lambda_{1}=0$. The inset in (a) shows the synchronization error
against the coupling constant for $N=7$. The orange horizontal
lines in (a) denote all eigenvalues except for $\lambda_{7}=2$ for
$N=7$. The orange lines and the white line in (b) denote the
eigenvalues except for the one characterizing the synchronous mode
for $N=4$ and $N=3$, respectively.}
\end{figure}

Now we consider a ring of Lorenz oscillators in which $N>2$. In a
ring structure, the eigenvalues of the coupling matrix
$\mathbf{C}$ take the form of $2\cos\frac{2i\pi}{N}$ and
distribute between -2 and 2. Larger $N$, more denser the
eigenvalues of $\mathbf{C}$. The stability of the synchronous
dynamics can be treated in a two-step procedure. In the first
step, the largest transversal Lyapunov exponent for each $\lambda$
is calculated using Eq.~(\ref{eq_4}) and Eq.~(\ref{eq_2}). Then
the region in the complex plane of $\lambda$, in which the largest
transversal Lyapunov exponent is positive, is obtained. In the
second step, the eigenvalues of the matrix $C$ are calculated. If
there is any eigenvalue except for the one for synchronous
dynamics falling onto the region, the synchronous dynamics is
unstable. Otherwise, the synchronous dynamics is stable. following
the procedure, the stability of the synchronous dynamics for a
ring of oscillators with the size $N$ can be determined. A
specific example is given in Fig.~\ref{fig_4}(a) where the
dependence of $\Lambda_{1}$ on $\lambda$ and $\epsilon$ is
presented with
$\mathcal{D}_{1}=\left(\begin{array}{ccc}0&0&0\\0&0&0\\1&0&0\end{array}\right)$
and
$\mathcal{D}_{2}=\left(\begin{array}{ccc}0&0&0\\0&0&0\\0&0&1\end{array}\right)$.
The black curve in the plot denotes the set of $\lambda$ and
$\epsilon$ at which $\Lambda_{1}=0$. Then we consider $N=7$. The
seven eigenvalues are plotted in Fig.~\ref{fig_4}(a) in orange
lines. To be noted, the eigenvalue $2\cos\frac{2i\pi}{N}$ is the
same as $2\cos\frac{2(N-i)\pi}{N}$ and there are only four orange
lines besides $\lambda_{N}=2$ in the plot. According to the
condition that the stability of the synchronous motion requires
$\Lambda_{1}<0$ for all transversal modes (or in another word, at
any $\epsilon$, the seven eigenvalues except for $\lambda=2$
should fall into the region with $\Lambda_{1}<0$), the figure
shows that there exists a large range of $\epsilon$ for stable
synchronous motions. As a comparison, we plot the synchronization
error against $\epsilon$ for $N=7$ in the inset, which shows a
disconnected regimes of $\epsilon$ for stable synchronous motion
and is in agreement with the analysis based on the dependence of
$\Lambda_{1}$ on $\lambda$ and $\epsilon$. In the case of
nonconjugate coupled oscillators, there is a size instability for
synchronous motions: At given oscillator parameters and coupling
constant, the synchronous state can only be realized when the
number of oscillators is below a threshold \cite{yang98}. However,
an extraordinary feature in Fig.~\ref{fig_4}(a) is that the
stability of the synchronous motion may be independent of the
number of oscillators in the ring in certain ranges of $\epsilon$.
For example, $\epsilon>0.25$ in which $\Lambda_{1}<0$ provided
that $\lambda\neq2$ and, consequently, the requirement of stable
synchronization is always satisfied regardless of $N$.

\begin{figure}
\includegraphics[width=3.4in]{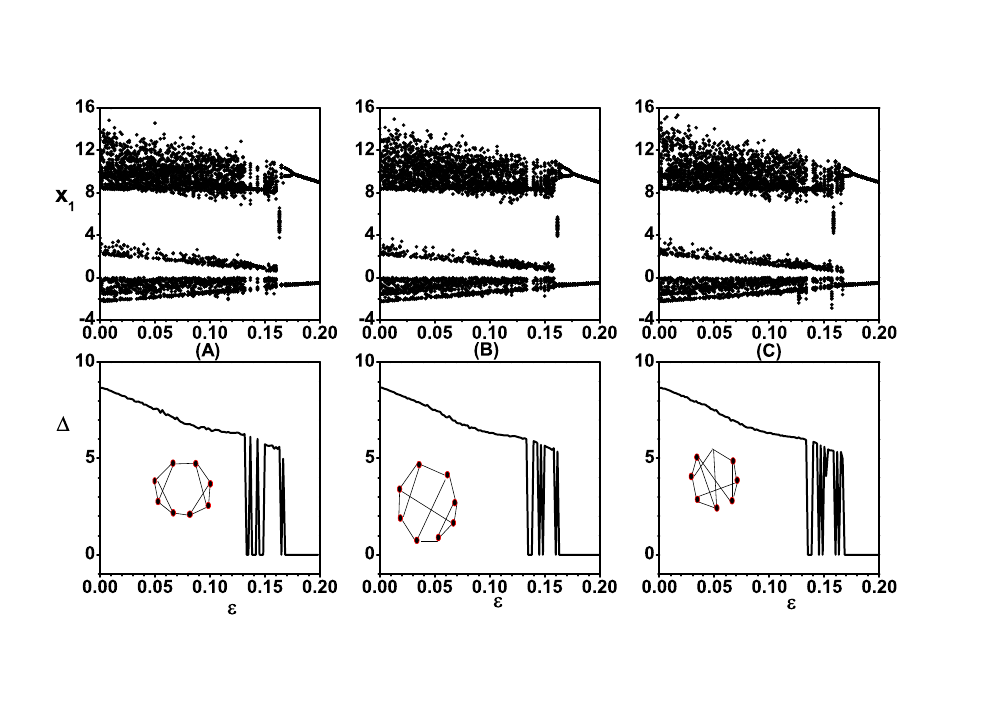}%
\caption{\label{fig_5} The dynamics of Lorenz oscillators with
conjugate coupling on different random networks. $\sigma=10$,
$r=28$, and $\beta=1$. The top panel shows the bifurcation diagram
of oscillators and the bottom shows the synchronization error. The
inset in the plots (A)-(C) in the bottom panel shows the
underlying network where each oscillator has three neighbors.
$N=8$, $\mathbf{D}_{1}$ and $\mathbf{D}_{2}$ are the same as those
in Fig.~\ref{fig_1}.}
\end{figure}

Then, we consider the case with
$\mathcal{D}_{1}=\left(\begin{array}{ccc}0&1&0\\0&0&0\\0&0&0\end{array}\right)$
and
$\mathcal{D}_{2}=\left(\begin{array}{ccc}1&0&0\\0&0&0\\0&0&0\end{array}\right)$,
which are the same as those in Fig.2. The results in
Fig.~\ref{fig_4}(b) shows that the synchronous motion becomes
independent of the number of oscillators when $\epsilon>12$. For
$\epsilon<12$, the synchronous motion can only be realized for
small number of oscillators, for example $N<4$. For $N=3$,
Fig.~\ref{fig_4}(b) also shows that stability of the synchronous
motion is non-monotonically dependent on $\epsilon$ since the
eigenvalues $\lambda_{1}=2\cos\frac{2\pi}{3}$ and
$\lambda_{2}=2\cos\frac{4\pi}{3}$ cross the black curves denoting
$\Lambda_{1}=0$ several times. Furthermore, we consider the case
with
$\mathcal{D}_{1}=\left(\begin{array}{ccc}0&1&0\\0&0&0\\0&0&0\end{array}\right)$
and
$\mathcal{D}_{2}=\left(\begin{array}{ccc}0&0&1\\0&0&0\\0&0&0\end{array}\right)$.
The synchronous motion realized in this case is a quenched state
which is not the solution of isolated oscillator. Figures
~\ref{fig_4}(c) and (d) show the results for increasing $\epsilon$
and decreasing $\epsilon$, respectively. The two figures show that
oscillators conjugate coupled in this way exhibit strong
hysteresis and there exists a range of $\epsilon$ in which
synchronous motion coexists with de-synchronous motions. Another
feature in these two figures is that the synchronous motion is
independent of the number of oscillators once it is realized.

The theoretical frame proposed here is not limited to ring
structures and it may be applicable to the regular random networks
in which each oscillator has the same number of neighbors. The
synchronous dynamics in a regular random network with degree $d$
follows
\begin{eqnarray}\label{eq_8}
\dot{\mathbf{s}}=\mathbf{f}(\mathbf{s})+d\epsilon
(\mathcal{D}_{2}-\mathcal{D}_{1})\mathbf{s}.
\end{eqnarray}
The stability of the synchronous dynamics on a regular random
network can be determined in the method presented
Fig.~\ref{fig_4}. The difference lies in that the range of
$\lambda$ to be concerned is dependent of $d$. We consider the
system consisting of eight Lorenz oscillators with the same $D_{1}$
and $D_{2}$ as those in Fig.~\ref{fig_1}. We presented the
bifurcation diagrams and the synchronization errors for the
oscillators on three regular random networks in Fig.~\ref{fig_5}.
Though oscillators are sitting on different networks, the
synchronous dynamics in these three cases is the same, which is in
agreement with the above analysis. As shown in Fig.~\ref{fig_5},
The discrepancies on the range of $\epsilon$ for stable
synchronous dynamics in these three cases are resulted from the
different eigenvalue spectrums realized by the underlying
networks.

In discussion, we have considered the synchronization in the
system of oscillators with conjugate coupling in which oscillators
interact through the coupling of dissimilar variables. We proposed
a general theoretical frame work for the synchronous dynamics and its
stability. We found that the synchronous dynamics and its
stability are dependent on both coupling scheme and the coupling
constant. We found that the stability of synchronous dynamics may
be independent of the number of oscillators, which is in contrast
to the size instability in the system of oscillators with
non-conjugate coupling. We also show that the theoretical analysis
in this work is applicable to regular random networks. However,
the synchronization among oscillators sitting in an arbitrary
complex network and with conjugate coupling pose a question which
worths further investigations.

\end{document}